\documentclass[11pt]{article}
\usepackage{moriond,epsfig}

\def\Journal#1#2#3#4{{#1} {\bf #2}, #3 (#4)}

\def\PRL{\em Phys. Rev. Lett.}
\def\PRD{{\em Phys. Rev.} D}
\def\PR{{\em Phys. Rev.}}

\def\be{\begin{equation}}
\def\ee{\end{equation}}
\def\bea{\begin{eqnarray}}
\def\eea{\end{eqnarray}}

\newcommand{\me}{\mathrm{e}}
\newcommand{\mi}{\mathrm{i}}

\newcommand{\mel}[3]%
    {\ensuremath{\left\langle #1 \left| #2 \right| #3 \right\rangle}}

\newcommand{\eplus}  {\ensuremath{e^+}}
\newcommand{\eminus} {\ensuremath{e^-}}

\newcommand{\pizero} {\ensuremath{\pi^0}}

\newcommand{\klong}  {\ensuremath{K_L}}

\newcommand{\gammas}  {\ensuremath{\gamma^*}}

\newcommand{\pizerod} {\ensuremath{\pizero_{{\rm D}}}}

\newcommand{\decay}[2]{\ensuremath{#1\to#2}}
\newcommand{\pigg}{\ensuremath{%
\decay{\pizero}{\gamma\gamma}}}
\newcommand{\pieeg}{\ensuremath{%
\decay{\pizero}{\eplus\eminus\gamma}}}
\newcommand{\pieeee}{\ensuremath{%
\decay{\pizero}{\eplus\eminus\eplus\eminus}}}
\newcommand{\pieeeeg}{\ensuremath{%
\decay{\pizero}{\eplus\eminus\eplus\eminus\gamma}}}
\newcommand{\klpipipi}{\ensuremath{%
\decay{\klong}{\pizero\pizero\pizero}}}
\newcommand{\klpipidpid}{\ensuremath{%
\decay{\klong}{\pizero\pizerod\pizerod}}}
\newcommand{\piggvert}{\ensuremath{\pizero\gammas\gammas}}

\newcommand{\ttwo}[2] {\ensuremath{g_{#1#2}}}
\newcommand{\tfour}[4]{\ensuremath{\epsilon_{#1#2#3#4}}}
\newcommand{\couplep} {\ensuremath{\tfour{\mu}{\nu}{\rho}{\sigma}}}
\newcommand{\couples} {\ensuremath{\left(\ttwo{\mu}{\rho}\ttwo{\nu}{\sigma}-\ttwo{\mu}{\sigma}\ttwo{\nu}{\rho}\right)}}

\newcommand{\units}[2]{\ensuremath{#1\,#2}}

\begin{document}
\vspace*{4cm}
\title{MEASUREMENTS FROM KTeV OF RARE DECAYS OF THE $K^0_L$ AND $\pi^0$}

\author{ E.\ D.\ ZIMMERMAN }

\address{University of Colorado\\
Boulder, Colorado 80302 USA}

\maketitle\abstracts{ The KTeV collaboration at Fermi National
  Accelerator Laboratory has recently completed searches for and
  measurements of several decay modes of the neutral kaon and
  pion. These include new searches for lepton flavor violating decays
  (which have not been seen), and a new study of the parity properties
  of the decay $\pi^0\rightarrow e^+e^-e^+e^-$.  }

\section{The KTeV Detector}

\begin{figure}
{\center
\psfig{figure=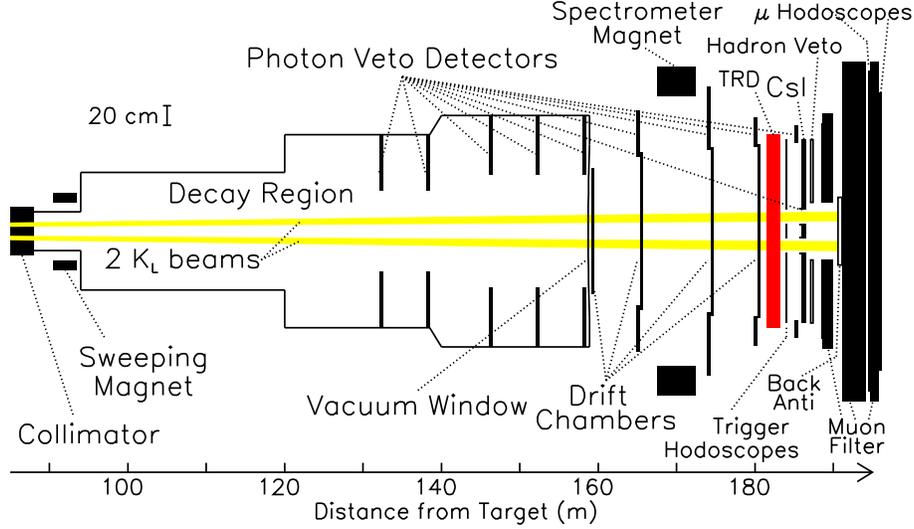,height=3in}\\}
\caption{The KTeV spectrometer as configured for E799.
\label{fig:spect}}
\end{figure}

Fermilab's KTeV detector (Fig.~\ref{fig:spect}) was constructed for
Experiments 799 and 832.  The two experiments were designed to
concentrate on different aspects of neutral kaon physics: E799 on rare
decays of the $K_L$ and E832 on measurement of ${\rm
  Re}(\epsilon^\prime / \epsilon)$.  A primary proton beam with energy
800~GeV struck a BeO target at a targeting angle of 4.8~mrad, and
collimation and sweeping magnets produced two parallel neutral hadron
beams.  The beams entered a 60~m long vacuum decay region, which ended
at a Mylar-Kevlar vacuum window.  Decay products were tracked with a
series of drift chambers surrounding a dipole analysis magnet.
Downstream of the drift chambers were a series of transition radiation
detectors (TRD) (in E799 only) and a pure CsI electromagnetic calorimeter, an
acive hadron beam absorber, and a set of muon detectors behind steel
shielding.  Photon veto detectors surrounded the fiducial volume in
the transverse directions.  The detector is described in more detail
in Ref.~\cite{ktev799}.

\section{The decay $\pi^0 \rightarrow e^+e^-e^+e^-$ and the 
  parity of the $\pi^0$}

The neutral pion's parity has historically been studied in two ways:
indirectly via the cross-section of $\pi^-$ capture on
deuterons~\cite{panofsky,chinowsky}, or directly via the double Dalitz
decay $\pi^0 \rightarrow e^+e^-e^+e^-$~\cite{samios}.  While both sets
of results are consistent with the negative parity, the direct
measurement has only 3.6$~\sigma$ significance.  KTeV has now reported
results \cite{pi04e-prl} that conclusively confirm the negative
$\pi^0$ parity as well as the first-ever searches for parity and $CPT$
violaton, and the first measurements of the electromagnetic form factor,
in this mode.

\begin{figure}
{\center
\psfig{figure=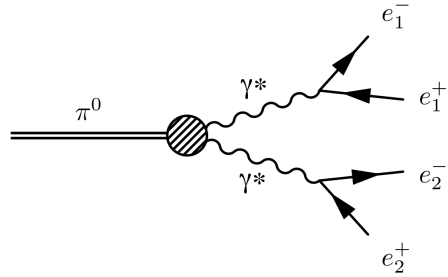,height=1.5in}\\}
\caption{Lowest order Feynman diagram for $\pi^0 \rightarrow e^+e^-e^+e^-$.
  The direct contribution is shown; a second diagram exists with
  $e_1^+$ and $e_2^+$ exchanged.
\label{ddfeyn}}
\end{figure}

The $\pi^0 \rightarrow e^+e^-e^+e^-$ decay proceeds through a two-photon
intermediate state (Fig.~\ref{ddfeyn}). 
The most general interaction Lagrangian for the $\pi^0\rightarrow\gamma^*\gamma^*$ transition
can be written \cite{barker}:
\begin{equation}
{\cal L} \propto C_{\mu\nu\rho\sigma}F^{\mu\nu}F^{\rho\sigma}\Phi
\end{equation}
where $F^{\mu\nu}$ and $F^{\rho\sigma}$ are the photon fields, $\Phi$
is the pion field, and the coupling has the form
\begin{equation}
C_{\mu\nu\rho\sigma} \propto  f(x_1, x_2)
      [ \cos\zeta\couplep \\*
      + \sin\zeta\me^{\mi\delta}\couples].
\end{equation}
The first term in $C_{\mu\nu\rho\sigma}$ is the expected pseudoscalar
coupling and the second term introduces a scalar coupling with a
mixing angle $\zeta$ and a phase difference $\delta$.  Nuclear parity
violation would introduce a nonzero $\zeta$, while $CPT$ violation
would cause the phase $\delta$ to be nonzero.  We assume the standard
parity-conserving form for the $\gamma^*\rightarrow e^+e^-$
conversion.

The form factor $f(x_1,x_2)$ is expressed in terms of the momentum
transfer of each of the virtual photons, or equivalently the invariant
masses of the two Dalitz pairs: $x_{1} \equiv
(m_{e^+_{1}e^-_{1}}/M_{\pi^0})^2;$ $x_{2} \equiv
(m_{e^+_{2}e^-_{2}}/M_{\pi^0})^2$.  In calculating the phase space
variables for an individual event, there is an intrinsic ambiguity in
assigning each electron to a positron to form a Dalitz pair.  KTeV's
analysis uses a matrix element model that includes the exchange
diagrams and therefore avoids the need to enforce a pairing choice.
The form factor is parametrized using a model
based on that of D'Ambrosio, Isidori, and Portol\'{e}s
(DIP)~\cite{dip}, but with an additional constraint that ensures the
coupling vanishes at large momenta~\cite{toalethesis}. In terms of the
remaining free parameters, the form factor is:
\begin{equation}
f_{\rm DIP}(x_1, x_2;\alpha) = \frac{1 - \mu (1+\alpha) (x_1+x_2)}
                                         {(1-\mu x_1)(1-\mu x_2)},
\end{equation}
where $\mu = M_{\pizero}^2/M_\rho^2 \approx 0.032$. 

The parity properties of the decay can be extracted from the angle
$\phi$ between the planes of the two Dalitz pairs in
Fig.~\ref{ddfeyn}, where pair 1 is defined as having the smaller
invariant mass.  The distribution of this angle from the
dominant direct contribution has the form $d \Gamma / d \phi \sim 1 -
A \cos (2\phi) + B \sin (2\phi)$, where $A\approx 0.2 \cos (2\zeta)$
and $B\approx 0.2\sin(2\zeta)\cos\delta$. A pure pseudoscalar
coupling, therefore, would produce a negative $\cos (2\phi)$
dependence.

The branching ratio measurement, which we describe here first, makes
use of a normalization mode in which two pions decay via $\pieeg$ and
the third $\pigg$. This ``double single-Dalitz'' mode, denoted
$\klpipidpid$ where $\pi^0_{\rm D}$ refers to $\pieeg$, has the same
final state particles as the signal mode.  Both modes are fully
reconstructed in the detector and the total invariant mass is
required to match the kaon's.  The two modes are distinguished by a $\chi^2$
formed of the three reconstructed $\pizero$ masses. This serves to
identify the best pairing of particles for a given decay hypothesis,
as well as to select the more likely hypothesis of the two.  The
similarity of these modes allows cancellation of most detector-related
systematic effects in the branching ratio measurement, but also allows
each mode to be a background to the other.

Radiative corrections complicate
the definition of the Dalitz decays in general.  We define the signal
mode $\pieeee$ to be inclusive of radiative final states where the
squared ratio of the invariant mass of the four electrons to the
neutral pion mass $x_{4e}\equiv (M_{4e}/M_{\pi^0})^2$ is greater than
$0.9$, while events with $x_{4e}<0.9$ (approximately 6\% of the total
rate) are treated as $\pieeeeg$.  
For normalization, the decay $\pieeg$ is
understood to include all radiative final states, for consistency with
previous measurements of this decay \cite{schardt}.  
Radiative corrections in this analysis are taken from an analytic
calculation to order $\mathcal{O}(\alpha^2)$ ~\cite{barker}.  

Radiative corrections complicate the definition of the Dalitz decays
in general.  The signal mode $\pieeee$ is defined to be inclusive of
radiative final states where the squared ratio of the invariant mass
of the four electrons to the neutral pion mass $x_{4e}\equiv
(M_{4e}/M_{\pi^0})^2$ is greater than $0.9$, while events with
$x_{4e}<0.9$ (approximately 6\% of the total rate) are treated as
$\pieeeeg$.  Radiative corrections in
this analysis are taken from an analytic calculation to order
$\mathcal{O}(\alpha^2)$ ~\cite{barker}.

The final event sample contains 30~511 signal candidates with
$\units{0.6}{\%}$ residual background and 141~251 normalization mode
candidates with $\units{0.5}{\%}$ background (determined from the
Monte Carlo simulation).  The background in the signal sample is
dominated by mistagged events from the normalization mode.
v
KTeV finds the following the ratio of decay rates: 
\begin{equation}
\frac{B_{eeee}^{ x>0.9} \cdot B_{\gamma\gamma}}{B^{2}_{ee\gamma}} = 
0.2245 \pm 0.0014{\rm (stat)} \pm 0.0009{\rm
  (syst)}.\end{equation} The $\pieeee$ branching ratio can
be calculated from the double ratio using the known values
$B_{\gamma\gamma}= 0.9980\pm 0.0003$ and $B_{ee\gamma}= (1.198\pm
0.032)\times 10^{-2}$ \cite{dalitzbr}.  This yields $B_{eeee}^{x>0.9}=
(3.26 \pm 0.18)\times 10^{-5}$, where the error is dominated by the
uncertainty in the $\pieeg$ branching ratio.  KTeV uses the radiative
corrections model \cite{barker} to extrapolate to all radiative final states, 
finding:
\begin{equation}
\frac{B_{eeee(\gamma)} \cdot B_{\gamma\gamma}}{B^{2}_{ee\gamma}} = 0.2383 \pm 0.0015{\rm (stat)} \pm 0.0010{\rm (syst)},\end{equation}
and $B_{eeee(\gamma)}=(3.46\pm 0.19)\times 10^{-5}$.
This branching ratio result is in good agreement with
previous measurements \cite{samios}.

The parameters of the $\piggvert$ coupling are found by maximizing
an unbinned likelihood function composed of the differential decay
rate in terms of ten phase-space variables.  The first five are $(x_1,
x_2, y_1, y_2, \phi)$, where $x_1$, $x_2$, and $\phi$ are described
above and the remaining variables $y_1$ and $y_2$ describe the energy
asymmetry between the electrons in each Dalitz pair in the $\pi^0$
center of mass \cite{barker}.  The remaining five are the same variables, but
calculated with the opposite choice of $e^+e^-$ pairings.  The
likelihood is calculated from the full matrix element including the
exchange diagrams and $\mathcal{O}(\alpha^2)$ radiative corrections.

The fit yields the DIP $\alpha$ parameter and the (complex) ratio of
the scalar to the pseudoscalar coupling.  For reasons of fit
performance, the parity properties are fit to the equivalent
parameters $\kappa$ and $\eta$, where $\kappa + \mi \eta \equiv \tan
\zeta \me^{\mi\delta}$. The shape of the minimum of the likelihood
function indicates that the three parameters $\alpha$, $\kappa$, and
$\eta$ are uncorrelated.  Acceptance-dependent effects are included as
a normalization factor calculated from Monte Carlo simulations.

Systematic error sources on $\alpha$ and $\kappa$ are similar to those
for the branching ratio measurement.  The dominant systematic error is due to
variation of cuts, resulting in a total systematic error of 0.9 and
0.011 on $\alpha$ and $\kappa$ respectively. For the $\eta$ parameter,
the primary uncertainty results from the resolution on the angle
$\phi$ between the two lepton pairs.%, which produces an effective
%flattening of the angular distribution without inducing a phase shift.
This behavior was studied
with Monte Carlo simulation and a correction was calculated. The
uncertainty on this correction results in a systematic error of 0.031.

\begin{figure}
{\center \includegraphics[width=3.4in]{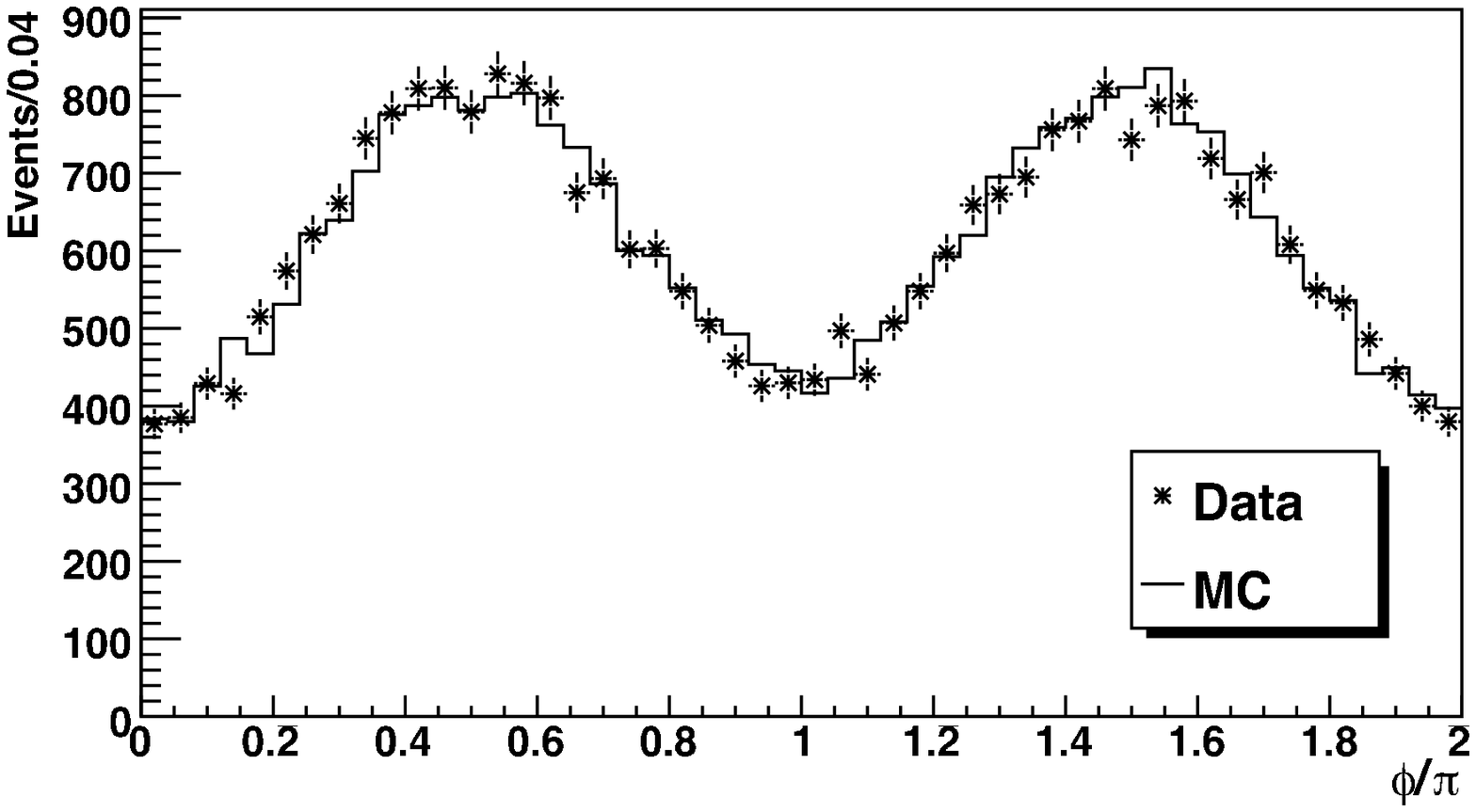}\\}
\caption{\label{phi} Distribution of the angle $\phi$, in units of
  $\pi$, between the planes of the two $e^+e^-$ pairs for $\pieeee$ candidate decays.  The solid
  histogram shows the Monte Carlo expectation for negative parity.}
\end{figure}

The $\phi$ distribution is shown in Fig.\ \ref{phi}.  For plotting the
data a unique pairing of the four electrons is chosen such that
$x_1<x_2$ and the product $x_1x_2$ is minimized: this choice
represents the dominant contribution to the matrix element.  It is
clear that the pseudoscalar coupling dominates, as expected, with no
evidence for a scalar component.  The distributions of all
five phase space variables agree well with the Monte Carlo simulation.

The parameters $\kappa$ and $\eta$
are transformed into limits on the pseudoscalar-scalar mixing angle
$\zeta$ under two hypotheses. If $CPT$ violation is allowed, then
the limit is set by the uncertainties in $\eta$, resulting in $\zeta <
6.9^\circ$ at the $90\%$ confidence level. If instead, $CPT$
conservation is enforced, $\eta$ must be zero, and the limit derives
from the uncertainties on $\kappa$, resulting in $\zeta < 1.9^\circ$,
at the same confidence level. These limits on $\zeta$ limit the
magnitude of the scalar component of the decay amplitude, relative to the
pseudoscalar component, to less than $12.1\%$ in the presence of $CPT$
violation, and less than $3.3\%$ if $CPT$ is assumed conserved.  The
limits on scalar contributions apply to all $\pi^0$ decays with 
two-photon intermediate or final states.

This analysis confirms the negative parity of the neutral pion with
much higher statistical significance than the previous result, and
places tight limits on nonstandard scalar and $CPT$-violating
contributions to the $\pieeee$ decay.  

\section{Lepton Flavor Violation}

Lepton Flavor Violation (LFV) in weak decays is a key signature of
several beyond-Standard Model physics scenarios.  Supersymmetry
\cite{ellis}, new massive gauge bosons \cite {landsberg,cahn}, and
technicolor \cite {technicolor} all can lead to LFV decays which might
be within reach of current experiments. Searches in $K_L$ decays are
complementary to searches in the charged lepton sector, since $K_L$
decays probe the $s \rightarrow d \mu e$ transition \cite{landsberg}.
KTeV-E799 has searched for the decays $K_L \rightarrow \pi^0 \mu^{\pm}
e^{\mp}$ and $\pi^0 \rightarrow \mu^{\pm} e^{\mp}$, and has made the
first reported search for $K_L \rightarrow \pi^0 \pi^0 \mu^{\pm}
e^{\mp}$ \cite{lfv799}.

In each case, the analysis required two charged tracks, one of which
was identified as a muon and the other an electron.  The key detector
elements for particle identification were $E/p$ in the CsI
calorimeter, response of the TRD, and muon hodoscopes downstream of
the muon filter steel.  Clusters in the CsI with no tracks pointing to
them were considered photons.

\subsection{$K_L \rightarrow \pi^0 \mu^{\pm} e^{\mp}$}

The dominant background for $K_L \rightarrow \pi^0 \mu^{\pm} e^{\mp}$
was the decay $K_L \rightarrow \pi^{\pm}e^{\mp}\nu_e$ ($K_{e3}$), with
a $\pi^{\pm}$ decay or punch through to the muon hodoscopes,
accompanied by two accidental photons faking a $\pi^0$. Since
accidental photons were often accompanied by other accidental
activity, we removed events with evidence of additional in-time
activity in the detector.  Additionally, the two photons were required
to form a good $\pi^0$ mass, and the square of the $\pi^0$ momentum in
the $K_L$ rest frame was required to be positive and therefore
physical.

\begin {figure}[h]
{\center \includegraphics[width=3.3in] {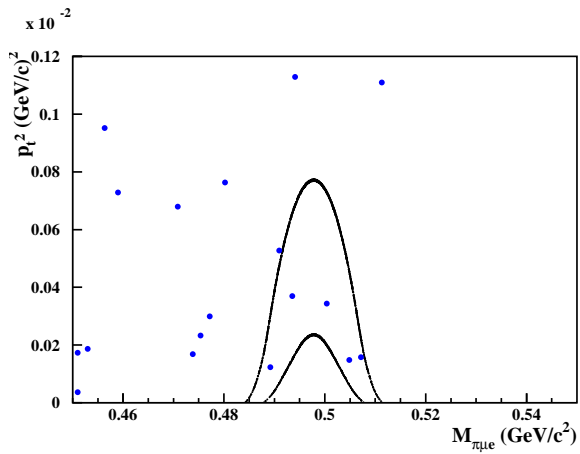}\\}
\caption{ Surviving events in the $p_t^2-M_{\pi^0 \mu e}$  plane for the 
$K_L \rightarrow \pi^0 \mu^{\pm}e^{\mp}$ search data. The signal and 
control regions are shown as the inner and outer solid contours. }
\label{result}
\end{figure}

The signal and control regions were defined using a likelihood
variable {\bf {\em L}} derived from $p_t^2$, the sum of the momentum
components of all final-state particles perpendicular to the kaon
flight line, and $M_{\pi^0 \mu e}$, the invariant mass of the $\pi^0
\mu e$ system.  The signal (control) region was defined by a cut on
{\bf {\em L}} chosen to retain 95\% (99\%) of signal Monte Carlo
events after all other cuts.  Expected background
levels were 0.66 $\pm 0.23$ events in the signal region and 4.21
$\pm 0.53$ events in the control region.  Both the signal and
control regions were blind during the analysis.  Figure
\ref{result} shows the $p_t^2-M_{\pi^0 \mu e}$ plane after all
cuts: five events were found in the control region and zero in the
signal.  The resulting limit is
 $B(K_L \rightarrow \pi^0 \mu^{\pm} 
e^{\mp}) < 7.56 \times 10^{-11}$ at 90\% CL, a factor 
of 82 improvement over the previous best 
limit for this mode. \cite{old1}

\subsection{Other lepton flavor violating modes}

KTeV has also searched for the decay $K_L \rightarrow \pi^0 \pi^0
\mu^{\pm} e^{\mp}$.  Reconstructing a second $\pi^0$ greatly reduces
the backgrounds, so some particle identification and anti-accidental
cuts were relaxed to improve the signal acceptance.  A similar analysis,
including a cut on a kinematic likelihood variable,
yielded no events in either the control region or the
signal region.  This resulted in a limit $B(K_L \rightarrow \pi^0 \pi^0
\mu^{\pm} e^{\mp}) < 1.64 \times 10^{-10}$.  This is the first limit
reported for this decay.

The decay chain $\klpipipi$, $\pi^0 \rightarrow \mu^\pm e ^\mp$ gives
the same final state particles as $K_L \rightarrow \pi^0 \pi^0
\mu^{\pm} e^{\mp}$, and therefore the same analysis procedure applies
with the additional requirement that the invariant mass $M_{\mu e}
\approx M_{\pi^0}$.  Since no events were found, the limit is
$B(\pi^0 \rightarrow \mu^{\pm} e^{\mp}) < 3.59 \times 10^{-10}$.  This
limit on $\pi^0 \rightarrow \mu^{\pm} e^{\mp}$ is equally sensitive to
both charge modes, while the previous best limits were not \cite
{zeller1}. Assuming equal contributions from both
charge combinations, KTeV's result is about a factor of two better than
the previous best limit on $\pi^0 \to \mu^+e^-$ and about a factor of
10 greater than the previous best limit on $\pi^0 \to \mu^-e^+$.

\section{Conclusion}
KTeV has completed several measurements recently on the decays of neutral
$K$ and $\pi$ mesons.  The measurement of $\pieeee$ represents the 
best direct determination of the parity of the $\pi^0$ and the first
searches for nonstandard parity and $CPT$ violation in this mode.  It
also yields the best
branching ratio and the first measurement of the form factor in this
mode. 
The limits on lepton flavor violation are now the
most stringent in the world for these decay modes.

\section*{Acknowledgments}
KTeV acknowledges the support and effort of the Fermilab
staff and the technical staffs of the institutions. This work was supported in part by the
U.S. DOE, NSF,
Ministry of Education and Science of Japan, Fundao de Amparo a
Pesquisa do Estado de S Paulo-FAPESP, Conselho Nacional de Desenvolvimento
Cientifico e Tecnologico-CNPq and CAPES-Ministerio Educao.

%\section*{Appendix}
% We can insert an appendix here and place equations so that they are
%given numbers such as Eq.~\ref{eq:app}.
%\be
%x = y.
%\label{eq:app}
%\ee

\end{document}